\documentstyle[prl,aps,epsf]{revtex}

\begin{document}
\bibliographystyle{prsty}
\twocolumn[\hsize\textwidth\columnwidth\hsize\csname@twocolumnfalse%
\endcsname
\draft
\title{Theory of Static and Dynamic Antiferromagnetic Vortices in LSCO Superconductors}
\author{Jiang-Ping Hu and Shou-Cheng Zhang}

\address{\it Department of Physics, McCullough Building, Stanford
  University, Stanford  CA~~94305-4045
}
\maketitle
\begin{abstract}
\end{abstract}
{\sl A key prediction of the $SO(5)$ theory is the antiferromagnetic vortex
state. Recent neutron scattering
experiment on LSCO superconductors revealed enhanced antiferromagnetic
order in the vortex state. Here we review theoretical progress since the
original proposal and present a theory of static and dynamic antiferromanetic
vortices in LSCO superconductors. It is shown that
the antiferromagnetic region induced by the vortices can be greater
than the coherence length, due to the light effective mass of the dynamic
antiferromagnetic fluctuations at optimal doping, and close promixity to the
antiferromagentic state in the underdoped regime. Systematic experiments are
proposed to unambiguously determine that the field induced magnetic scattering
originates from the vortices and not from the bulk.}

\pacs{ {\bf Invited Talk at SNS 2001}} ]

$SO(5)$ theory is an unified theory of antiferromagnetism (AF) and
superconductivity (SC) in the cuprate superconductors\cite{so5,pso5}.
Although AF and SC phases seem to be detached in the experimental
phase diagram, this theory predicts a direct transition between
them as the chemical potential is varied. As a function of a
physical parameter, the direction of the $SO(5)$ superspin vector
can be rotated smoothly. However, the chemical potential is hard to control
experimentally. Doping in the AF/SC transition region is complicated by the
chemical and structural inhomogeneities. Therefore, in the
$SO(5)$ theory, it was proposed that the interplay
between these two phases can be investigated in a controlled way
in the vortex state of the superconductor, where the external magnetic field
can smoothly rotate the direction of the superspin vector. This theory makes the
striking prediction that the core of the vortex is
antiferromagnetic\cite{so5}, in sharp contrast to the metallic vortex core
in a conventional superconductor. The AF vortex state (AFVS) can be viewed as
a topologically non-trivial texture of the $SO(5)$ superspin vector.
Furthermore, it was explicitly proposed that
the AF vortex state can be detected in neutron scattering experiments as
satellite peaks spaced by $2\pi/d$, where $d$ is the spacing
between the vortex cores\cite{so5}. These satellite peaks give a
precise experimental definition of the AF vortex state,
where static AF order coexists with the vortex state of the superconductor.

A theory of the AF vortex core was developed by Arovas et al\cite{arovas}.
This work introduced a method to study the vortex core based on
a Schroedinger like equation for the magnetic excitations near the
vortex core. Due to the $SO(5)$ constraint on the AF/SC superspin vector\cite{so5},
the vortex core acts like an attractive potential
for the magnetic excitations. If a bound state exists,
enhanced low energy magnetic fluctuation is localized near the
vortex core. If the bound state energy goes to zero, static AF order
is established inside the vortex core.
Based on estimation of parameters, it was argued that the static
vortex core is stable in the underdoped regime\cite{arovas}.
These authors suggested that the AF vortex state can be
detected in neutron scattering experiments, and predicted that
the field induced
antiferromagnetic moments should scale with the applied field, or the
number of vortices in the system\cite{arovas}. Soon after this work,
Bruus et al\cite{bruus} analyzed experimental data in YBCO superconductors
and argued that the vortex core in optimally doped materials should
have enhanced dynamic AF order, with energy below the neutron resonance
energy of $41 meV$. Ogata\cite{ogata} performed extensive variational
calculations on the $t-J$ model and found direct evidence of the static
AF vortex core up to doping level of $\delta=0.1$. In a classical Monte
Carlo calculation of the $SO(5)$ model, Hu\cite{hu} observed the AF vortex
state in the satellite peaks of the AF spin correlation function, which
are spaced by $2\pi/d$. Thermodynamic implications of the AF vortex state
has been recently investigated by Juneau et al\cite{juneau}.
Various groups\cite{andersen,ogata,lee,franz}
argued that the AF vortex core is expected to
lead to strong suppression of the local density
of states, consistent with the STM experiments\cite{stm}.
Recently, Demler et al\cite{demler} described the influence of the
superflow kinetic energy and found that this led to
an important contribution to the magnetic
field dependence of the spontaneous moment in a phase
with magnetic long-range order. Furthermore, these authors also
developed a theory for dynamic magnetism in the vortex lattice.
This approach is reviewed in the same volume by Sachdev\cite{sachdev_sns}.

On the experimental side, the search of AF vortex state showed promising
results. Early on Vaknin et al\cite{vaknin} found evidence of AF order in the
vortex state of YBCO. Katano et al\cite{yamada} found enhanced AF fluctuations
in the vortex state of LSCO. Most striking evidence of AF order in the vortex
state is observed recently by Lake et al\cite{lake1,lake2}.
In the vortex state of optimally doped LSCO superconductors,
incommensurate magnetic fluctuations were found in the low energy region
inside the spin gap, whose intensity scales with the number of vortices
in the system. In the underdoped LSCO superconductors, enhanced static AF
order was discovered in the vortex state. In another related system,
$La_2CuO_{4+y}$\cite{wells,ylee}, similar field induced enhancement has been discovered.

The purpose of this paper is to show that the theory of the AF vortex
state\cite{so5,arovas} can be applied straightforwardly to explain the
recent neutron scattering measurements by Lake et al\cite{lake1,lake2}.
We discuss various energy and length scales in the problem, and
show that experiments are performed in a regime where static and dynamic
AF vortex state can be observed. In the original theory, only the case
of the commensurate AF fluctuations were discussed. As we shall see, the
theory can be extended in a simple way to accommodate incommensurate
AF fluctuations as well. Most importantly, we show that one can determine
unambiguously whether the field enhanced AF fluctuation {\it originates} from the
bulk or from the vortices, and show that present experiments already provides
very strong evidence that the later case is realized.

The starting point of our consideration is the effective $SO(5)$ theory of the
magnetic fluctuations linearized in an inhomogeneous background of the SC
condensate\cite{so5,arovas}. This theory can be cast into the following form
of the effective Lagrangian ${\cal L}= \frac{1}{2\Delta_s}(\partial_t m_\alpha)^2 -H$,
where $m_\alpha(x,t)$ is the local AF order parameter and
\begin{eqnarray}
H=\frac{1}{2}\sum_p \Lambda_p m_\alpha(p) m_\alpha(-p) + \frac{1}{2}\sum_x v(x) m_\alpha(x)^2
\label{H}
\end{eqnarray}
where $\omega=\sqrt{\Lambda_p \Delta_s}$ is the dispersion of the AF fluctuation
in the absence of the external magnetic field. In the LSCO superconductors, we
can approximate it with the following form:
\begin{eqnarray}
\Lambda_p = \Delta_s + \frac{(p-Q_0)^2}{2M^*}
\label{omega}
\end{eqnarray}
Here $\Delta_s$ is the spin gap energy. In optimally doped LSCO superconductors,
$\Delta_s \approx 7 meV$. $Q_0$ is the wave vector of the incommensurate
AF fluctuations in LSCO\cite{ic}, which can be interpreted as
dynamic stripes\cite{zaanen,kivelson}. $M^*$ is the effective mass for the AF
fluctuations, and we shall give an estimate of this parameter later.
$v(x)$ is the potential due to the SC order parameter. In the $SO(5)$ theory,
there is a constraint between the AF and SC order parameters in the form
$m_\alpha^2 + |\psi|^2=1$, where $\psi$ is the SC order parameter.
If we implement this constraint by a soft spin
constraint, of the form $-g (m_\alpha^2 + |\psi|^2) + u (m_\alpha^2 + |\psi|^2)^2$,
we see that a repulsive coupling between the two order parameters is implied.
This would remain true in the presence of a $SO(5)$ symmetry breaking term
$w m_\alpha^2 |\psi|^2$, as long as $w>-2u$. The repulsion between the SC and the
AF order parameters is the crucial physics for all field induced magnetic
phenomena. If these two order parameters were decoupled\cite{aharony}, no
such effects would exist. Linearizing with respect to this
repulsive coupling in the bulk we obtain the spin gap $\Delta_s$. However,
at the center of the vortex core, the SC order parameter vanishes, and the effective
spin gap is lowered. Therefore, there is an effective attractive potenial $v(x)$
for the AF magnetic fluctuations inside the vortex core. We assume that it takes
the form
\begin{eqnarray}
v(x)=\sum_{i} v_0(x,x_i), \ \
v_0(x,x_i)=-V e^{-\frac{(x-x_i)^2}{2\xi^2}}
\label{v}
\end{eqnarray}
where $V<\Delta_s$, $\xi$ is the SC coherence length and $x_i$ describes the
center of the vortices. Equations (\ref{H}), (\ref{omega}) and (\ref{v}) determines
the AF fluctuation spectrum in the vortex state.

The description of the free propagation of the AF magnetic fluctuation obtained
from the linearized $SO(5)$ theory is similar to that first obtained from analytically
continuing a magnon triplet from a quantum disordered state, such as the spin
Peierls state\cite{sachdev,sachdev2}. However, the crucial point here is the attractive
interaction between the AF magnetic fluctuations with the vortices, which is
first obtained from the $SO(5)$ theory. In the original theory, a commensurate
form of the dispersion is assumed. However, it was later found that competing
interactions in the $SO(5)$ theory could produce dispersion in the form
of (\ref{omega}), with a general incommensurate wave vector $Q_0$\cite{jphu}.
For our current discussion, it is easy to see that $Q_0$ can be trivially gauged
away from the problem. If we perform a phase transformation
\begin{eqnarray}
m_\alpha(x) = e^{iQ_0 x} n_\alpha(x)
\label{gauge}
\end{eqnarray}
we see easily that the quadratic Hamiltonian can be diagonalized by the
following Schroedinger like equation for $n_\alpha(x)$:
\begin{eqnarray}
(-\frac{\nabla^2}{M^*}+v(x)) n_\alpha(x) = \lambda n_\alpha(x)
\label{Schroedinger}
\end{eqnarray}
with energy given by $\Delta_s+\lambda$. This equation was first
introduced in equation (10) of ref. \cite{arovas}, and forms the basis
for subsequent studies on AF order in the vortex state.

Equation (\ref{Schroedinger}) describes the motion of an effective
quantum particle with mass $M^*/2$ in a periodic array of attractive vortex
centers. The important physics here is that in D=2, at least one bound state
forms inside the attractive potential. This bound state for the collective AF fluctuation
is not to be confused with the bound state of quasi-particles inside the
vortex cores. Let us first consider the simple case where only one bound state
forms inside the attractive potential, with energy $E_B$. The localization
length of the bound state can be estimated to be
\begin{eqnarray}
l \sim \sqrt{\frac{\hbar^2}{M^* (\Delta_s-E_B)}}
\label{l}
\end{eqnarray}
In the limit where $l<d$, the band width of the vortex band can be
estimated to be
\begin{eqnarray}
t \sim V e^{-d/l}
\label{t}
\end{eqnarray}
The three length scales, the SC coherence $\xi$, the AF magnetic localization
length $l$, the inter-vortex spacing $d$, and the four energy scales,
the spin gap $\Delta_s$, the attractive potential $V$, the bound state energy
$E_B$ and the vortex band width $t$ are illustrated in Fig. 1.
The concept of a static or
dynamic AF vortex core is a sharply defined concept. If there is no bound state,
then there is no sense in which one can define the concept of AF vortex core.
However, if bound states exists inside the spin gap, the vortex core develops
dynamic AF ordering, with well defined ordering energy scale $\Delta_s-E_B$ and
well defined ordering length $l$. One can view $l$ as the size of the vortex
and due to the finite size quantization, the AF moment fluctuates dynamically,
with time scale $\hbar/(\Delta_s-E_B)$. In the regime of weak field, $l<<d$, the
bandwidth of the AF vortex band can basically be neglected. With increasing
field, the bound state energy $E_B$ decreases, and the inter-vortex coherence
$t$ increases as well. When the bottom of the vortex band energy touches zero,
the static AF vortex state is obtained.

With this basic preparation we can now discuss the experimental situation in
the optimally doped LSCO superconductors, and present the central argument
of this paper that the field induced scattering {\it originates} from
AF fluctuations localized around the vortex
cores. Experiments are done in the weak
magnetic field regime where the volume fraction associated with the vortices
is extremely small, $f \sim (\xi/d)^2 < 10 \%$. On the other hand, the field
induced magnetic scattering is centered around $4 meV$, well below the spin gap
energy of $\Delta_s = 7 meV$. If the field induced
AF magnetic fluctuations were extended, scattering by the vortices can only cause
an energy shift proportional to $V (\xi/d)^2$, which vanishes in the limit of low vortex
volume density. See Fig. \ref{fig2a}. In the experimental regime of $f \sim 10\%$, the
energy shift of
extended states would be of the order of $\Delta E=V f<\Delta_s f < 0.7 meV$,
far less than the experimentally observed value of $\Delta E\sim 3 meV$.
Therefore, the Lake et al experiment\cite{lake1} already shows that
it is extremely unlikely that the field induced scattering originates from the
bulk. On the other hand, if the field induced AF magnetic fluctuations are localized
around the vortices, the energy shift is finite in the limit $f \rightarrow 0$.
See Fig. \ref{fig2b}.
If we assume one bound state per vortex, the field induced intensity $I$ is proportional
to $f$. {\it Therefore, this picture predicts that the field induced intensity is directly
proportional to $H/H_{c2}$}. These are the precise signatures of the AF vortex core in the limit
of low vortex volume density. In Fig. 2 of ref. \cite{lake1},
one sees clearly that the continuum above the spin gap is little changed by the
field, but new spectral weight is introduced with a centroid well detached
from the continuum. The experimental fact that the centroid of the field induced
signal is well below the spin gap and the intensity is proportional to $f$ provides
strong evidence that the field induced scattering is localized around the
vortices. A more systematic analysis should plot both the energy centroid and
the intensity of the field induced signal as a function of $f$. Since $t$
depends exponentially on $d$, it vanishes faster than any other energy scales
in the limit of $f\rightarrow 0$.
{\it Therefore, if $E_B \rightarrow$ finite and $I \propto f$
in the extrapolated limit of $f\rightarrow 0$, unambiguous evidence in support
of the AF vortex core can be established.} The $f\rightarrow 0$
limit is an useful tool to label the bands, since once $f$ becomes finite, there is no
rigorous distinction between the extended and localized bands.

Considerations of the energy of the field induced scattering leads
to the conclusion that the they must be localized near the vortex
cores. Let us now fix the energy at $\Delta_s-E_B$ and consider
the spatial distribution of the AF fluctuations. The three length
scales can be easily estimated from the experiments, giving
$\xi\sim 4a_0$, $l\sim 20 a_0$ and $d\sim 40 a_0$. Here $l$ is
estimated from the width of the momentum distribution function at
fixed energy of $4 meV$. Theory predicts\cite{so5,hu,demler} that
the momentum distribution
function for the $n(x)$ field consists of a series of satellite
peaks spaced by $G=2\pi/d$, centered around the main peak at $q=0$,
where the intensity of each satellite peak is covered by an
envelope function of the width $1/l$. See figure (\ref{fig3}).
Neutron scattering couple to the $m(x)$ field, by applying the
transformation (\ref{gauge}) we obtain the same pattern, but
with the center of the envelope function shifted to incommensurate
wave vector $Q_0$. See figure (\ref{fig4}). {\it Since the momentum
shift, or gauge, transformation (\ref{gauge}) applies equally
to both extended and localized states, the AF fluctuation localized
near the vortex core is associated with the exactly the same incommensurate
wave vector $Q_0$ as the AF fluctuations in the bulk.}

Now we need a quantitative argument on why the observed AF coherence
length $l \sim 20 a_0$ seems longer compared to $\xi$. From equation
(\ref{l}), we see that this could be due to the light effective mass
$M^*$. But $M^*$ could be independently measured in the neutron
scattering experiment in the absence of the applied field, by
fitting the experimentally measured AF dispersion with
$\omega_p=\sqrt{\Lambda_p\Delta_s}$
given in equation (\ref{omega}). This provides an independent consistency
check on our theory. A small effective mass translates into steep
dispersion in $\omega_p$ versus $p$. Experimentally, the incommensurate
peak of the momentum distribution function for various energies does
seem to shift, implying a very small effective mass. If we assume that
the incommensurate peak does not shift over an energy range of $8 meV$,
and momentum shift less than $(10a_0)^{-1}$ can not be detected,
we obtain a value for $M^*$ which is roughly consistent with
the coherence length $l\sim 20 a_0$. It would be highly
desirable to improve the experimental accuracy and quantitatively check
the relation between the inelastic neutron scattering in the absence of the
field, bound state energy and the coherence length in the vortex state.

As the external magnetic field increases, $d$ decreases and the intervortex
coherence $t$ increases. For $d \sim l$, it ceases to take the form $t\sim V e^{-d/l}$.
A similar effect arises when the
system is more underdoped, where both $\xi$ and $l$ increases, resulting in
the increase of $t$. The local bound state near each vortex now form a band,
with band width $t$. As a function of the applied field, the center
of the band shifts downwards, and the bandwidth increases. Within Born scattering,
the shift of the band minimum can be estimated to be
\begin{eqnarray}
E_{min}= E_B - \alpha V f
\label{min}
\end{eqnarray}
where $\alpha$ is a dimensionless constant.
When the bottom of this vortex band touches zero, the system
makes a phase transition into a AF vortex state as predicted in\cite{so5,arovas}.
However, Demler et al \cite{demler} observed that in this regime, the
superflow makes an important correction to the field dependence, and find
a logarithmic correction\cite{lake2,ylee}.  As argued in previous
paragraphs, the static AF moments are not localized in a region of the size of
$\xi$, but of the size of $l$, which is now comparable to $d$.
It is important to point out here that due to the repulsion between
AF and SC order parameters, the static AF order near the vortices is always larger than
the AF order in the bulk. However, these
quantitative corrections do not change the basic periodic structure. As pointed
out in ref.\cite{so5,hu,demler}, this periodic pattern can be detected directly in neutron
scattering experiments as satellite peaks. One should observe the momentum distribution
functions displayed in Fig. 2 and 3, for quasi-elastic scatterings. These satellite
peaks provide direct experimental evidence for the static AF order in the vortex
state. Recent neutron scattering experiments in underdoped LSCO superconductors
revealed long ranged static incommensurate AF order in the presence of an external
magnetic field\cite{lake2,ylee}, where the field induced signal is proportional to $H/H_{c2}$,
or the number of vortices in the sample. It is highly desirable to refine the
momentum resolution in the experiments and detect the satellite peaks directly.

Now we are in a position to understand why relatively moderate field $H_{AFVS}$
can induce a transition into a statically ordered AF vortex state.
If there were no localized states near the vortices in the $f\rightarrow 0$ limit,
the transition to a statically ordered state arises from
the band of extended states, whose bottom touches zero. However, if the band of
localized state exists below the extended band, its bottom will always touch
zero first. Since bands are not expected to cross as the external magnetic
field is varied, we can roughly estimate the ratio between these two critical
fields to be $H_{AFVS}(bound)/H_{AFVS}(extended)\sim (\Delta_s-E_B)/\Delta_s\sim 1/2<1$.
In other words, transition due to the localized band always preempts the transition
due to the extended band. From this argument, we reach the important conclusion
that the {\it relatively moderate field can induce a transition into a statically
ordered AF vortex state precisely because the AF order originates from the vortices.}

In above discussions we mostly focused on neutron scattering experiments in LSCO.
However, the theory is generally applicable to all high $T_c$ materials. In recent
NMR experiments\cite{halperin}, enhanced AF fluctuations around the vortex cores
have been detected in the YBCO system. In a recent STM experiment,
Hoffman et al\cite{davis} detected CDW-like order around the vortex core in the
BSCO system. In both cases, the magnetic field induced order are well localized around
the vortex core, with a localization length greater than the vortex core itself.
The important point to be stressed here is that the localization length is finite,
in accordance to the general discussions outlined above. It would be highly
desirable to extend the neutron scattering experiments under the magnetic field
already performed by Dai et al\cite{dai} on YBCO systems to higher field, and
to look for the dynamic AF bound state around the vortex core, with bound state
energy of few $meV$ below the neutron resonance energy.

Finally, we would also like to comment on a new order parameter which emerges in the
AF vortex state. In ref.\cite{so5}, one of us introduced a general constraint
between the $SO(5)$ superspin order parameters $n_a$ and the $SO(5)$ symmetry
generators $L_{ab}$, which takes the form:
\begin{eqnarray}
\epsilon^{abcde} n_c L_{de}=0
\label{orthogonality}
\end{eqnarray}
In regions of space where both the AF order parameter $n_{2,3,4}$ and the
SC order parameters $n_{1,5}$ coexist, the $SO(5)$ constraint (\ref{orthogonality})
automatically implies that the corresponding $\pi$ operators $L_{12,13,14}$ and
$L_{25,35,45}$ would acquire non-zero expectation values as well. Therefore,
in the region $\xi<r<l$ around the vortex core, where both AF and SC orders
coexist, the order associated with the $\pi$ operators in the $SO(5)$ theory
exists as well. In a subsequent paper, we shall
fully explore the experimental consequence of this observation.

In conclusion we have reviewed theoretical progress since the original proposal
of AF order in the SC vortices\cite{so5,arovas}, and presented a semi-quantitative
theory of the static and dynamic AF order in the vortex state of LSCO superconductors.
It is argued that the field induced signals
found in experiments arise from the dyanmic and static AF order in the
superconducting vortices. We showed that the intermediate length scale $l$
describes the localization of AF order near SC vortices. This
length scale is large in the LSCO materials because the effective mass for the
incommensurate AF fluctuations in the absence of the field is anomalously
small. However, this length scale is finite in the limit of low vortex density.
Whether the field induced scattering originates from the bulk or from the vortices
can be systematically distinguished in neutron
scattering experiments, where the energy shift of the field induced scattering
is plotted against $H$, in the limit of low $H$. If the energy shift remains finite
in the zero field limit, an unambiguous case for the AF vortices can be established.
Satellite peaks are predicted in the momentum distribution functions for both
quasi-elastic and inelastic scatterings, revealing the static and dynamic AF order
of vortices. An exact argument based on gauge transformation is presented to show
that the AF order localized around the vortices has the same incommensurate
wave vector as the bulk AF fluctuations.

The theoretical prediction and experimental discovery of the AF vortex state
offers crucial
insights into the microscopic mechanism of high $T_c$ superconductivity.
Various theoretical proposals differ only in the nature of the intermediate state
between the AF and the SC state.
If doping or chemical potential were the only route connecting these two phases,
the problem of chemical inhomogeneities will always obscure the underlying
physics. The external magnetic field provides a clean alternate route to connect these
two phases and a new route to attack this
problem. Unlike the case of chemical doping, experiments, coupled with the scaling
arguments presented in this work, can precisely determine the real space locations of the
AF signals. While the $SO(5)$ theory predicts an AF vortex state\cite{so5},
other theories predict $hc/e$ vortex\cite{half}, visons\cite{vison} and
staggered flux phase\cite{staggered}.
Therefore, the nature of the vortex state is important in distinguishing among the various
theoretical proposals.

We would like to acknowledge useful discussions with Drs. G. Aeppli, S. Davis,
E. Demler, W. Hanke, W. Halperin, S. Kivelson, Y. Lee and S. Sachdev for useful discussions.
This work is supported by the NSF under grant numbers DMR-9814289.
JP Hu is also supported by the Stanford Graduate fellowship.

\bibliographystyle{unsrt}
\bibliography{sns01}

\begin{figure*}[h]
\centerline{\epsfysize=5.0cm \epsfbox{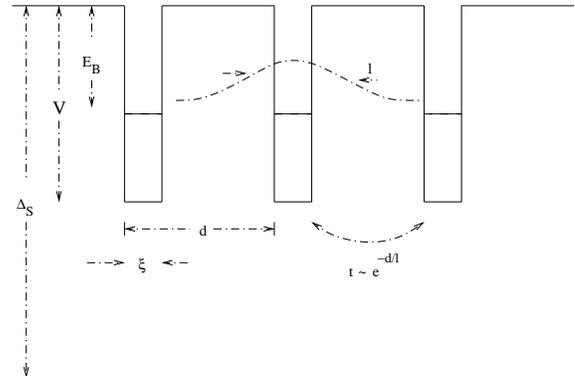} } \caption{
Illustration of the energy and length scales in the problem. $\Delta_s$
is the spin gap energy, $V$ is the attractive potential due to SC vortices,
$E_B$ is the bound state energy and $t$ is the
bandwidth of the AF vortex band.
$\xi$ is the SC coherence length, $l$ the localization length
of AF fluctuations around the vortices and $d$ is the inter-vortex
spacing. }
 \label{fig1}
\end{figure*}

\begin{figure*}[h]
\centerline{\epsfysize=5.0cm \epsfbox{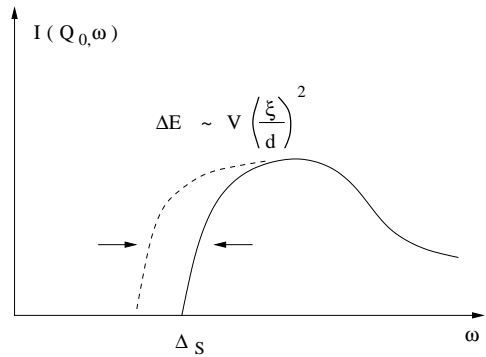} } \caption{
If no bound states exist, energy of extended states can only
shift by an amount of the order of $V(\xi/d)^2$, which is vanishingly
small in the limit of $d\rightarrow 0$.
}
\label{fig2a}
\end{figure*}

\begin{figure*}[h]
\centerline{\epsfysize=5.0cm \epsfbox{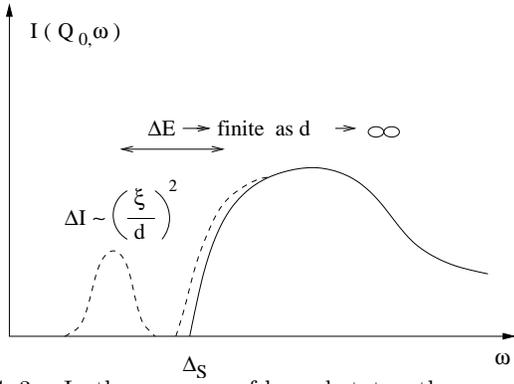} } \caption{
In the presence of bound states, the energy shift is finite, while
the intensity due to the bound state band is proportional to
$(\xi/d)^2\sim H/H_{c2}$.
}
\label{fig2b}
\end{figure*}

\begin{figure*}[h] \centerline{\epsfysize=5.0cm
\epsfbox{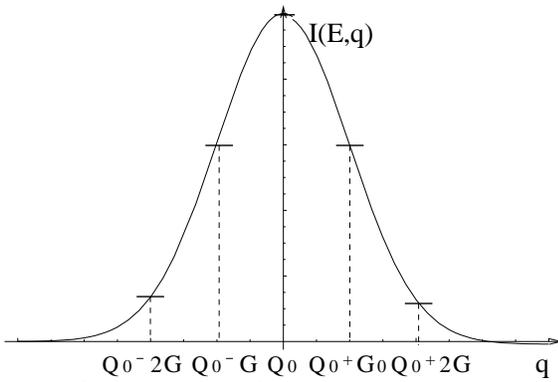} } \caption{At energy $E=\Delta_s-E_B$ for optimal
doping and $E=0$ for underdoping, the momentum distribution function
consists of a series a delta function peaks, spaced by $G=2\pi/d$. The
envelope of these peaks forms a broader peak centered around the incommensurate
wave vector $Q_0$, with width $1/l$.
}
\label{fig3}
\end{figure*}

\begin{figure*}[h] \centerline{\epsfysize=5.0cm
\epsfbox{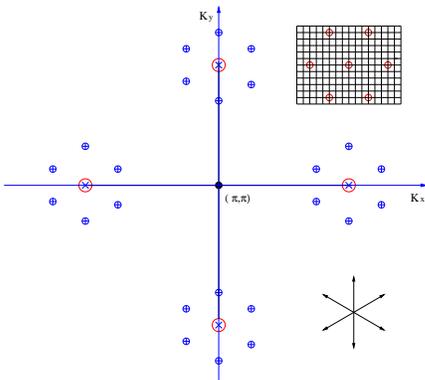} } \caption{A more explicit illustration of the
satellite peaks near the four incommensurate spots. The inset shows the
vortex lattice in real space.}
\label{fig4}
\end{figure*}

\end{document}